\documentstyle[aps,prd]{revtex}

\begin{document}
\title{Slowly Rotating Boson-Fermion Star}
\author{Claudio M. G. de Sousa (a,b)
\thanks{Electronic-addresses: claudio@unb.br, claudio@iccmp.br, desousa@ucb.br},
Vanda Silveira(a)\thanks{%
Electronic-address: vandalrs@tcu.gov.br} \\
{\em (a)-International Centre of Condensed Matter Physics,
Universidade de Bras\'{\i }lia, 
Caixa Postal 04513, 70919-970, Bras\'\i lia-DF, Brazil. \\
(b)-Diretoria de Matem\'{a}tica, Universidade Cat\'{o}lica de Brasilia,
QS 07, Lote 01, EPCT, 72022-900, Bras\'{i}lia-DF, Brazil. }
}

\maketitle

\begin{abstract}
{\bf Abstract: }Relativistic prescription is used to study the slow rotation
of stars composed by self-gravitating bosons and fermions\ (fermions may be
considered as neutrons). Previous results demand that purely boson stars are
unable to display slow rotation, if one uses relativistic prescription with
classical scalar fields. In contrast to this, the present work shows that a
combined boson-fermion star in its ground-state can rotate. Their structure
and stability are analysed under slow rotation approximations.
\end{abstract}

\section{Introduction}

Stars composed by neutrons (fermions) has been first suggested by
Oppenheimer and Volkoff \cite{Oppenheimer39}. After that, many theoretical
studies have arisen on this theme and some years later many stellar 
objects has been identified with neutron stars.
Inside neutron stars, fermions are attracted each other
gravitationally but the configuration is prevented from collapsing via Pauli
exclusion principle pressure.

Another possibility for stars configurations has been suggested \cite
{Ruffini69} independently by Kaup, and Ruffini and Bonazzola. Boson stars
are compact configurations made with bosons gravitationally bounded. Bosons
are described via a scalar field, and large interest has been shown on this
subject since properties of that kind of configurations can be obtained
directly from the Lagrangian treatment, without the need to define an
equation of state, as it is usually done to ordinary stars (see also Breit
et al \cite{Breit84}). Boson stars are prevented from collapsing
gravitationally due to a pressure that arises from the Heisenberg
uncertainty principle. Boson stars formation by Jeans instability has been
studied, {\em e.g.}, by Grasso \cite{Grasso90} and its stability has been
addressed by Gleiser \cite{Gleiser88} and by Torres \cite{Torres97}. Some
reviews \cite{Liddle92} present the relativistic and astrophysical interest
on this subject,  properties and detection expectations.

Since in the primordial gas possibly bosons and fermions coexisted, one
could expect stable configurations of self-gravitating bosons and fermions
formed. A model boson fermion star has been first proposed by Henriques,
Liddle and Moorhouse \cite{Henriques89}. Some studies on stability \cite
{Henriques90} and on configurations with interactions between bosons and
fermions \cite{Sousa98} has been performed.

Boson star rotation has been first addressed by Kobayashi {\em et al} \cite
{Kobayashi94}. They used a perturbative approach and have shown that boson
stars could not display slow rotation. But, as shown by Silveira and de
Sousa \cite{Silveira95}, considering the scalar field quantum nature with
axial symmetry it is possible to obtain stable configurations with $l\neq 0$%
, using Newtonian approach. After that, many works have contributed to the
comprehension of the properties of spinning boson stars, as Ryan \cite
{Ryan97} which shows the large self-interaction case, and Yoshida\ and
Eriguchi \cite{Yoshida97} studied static relativistic axisymmetric solutions.

But it seems there is no explicit proof that boson fermion stars can display
rotation. Hence, in this article we present our results for relativistic
slowly rotating stars composed with bosons and fermions. In section II we
present the construction of a boson star in II.A and of a boson fermion star
in section II.B. In section III we present the construction of a rotating
fermion star in III.A, and an extension for studying rotating
boson fermion stars in III.B. Results are shown in section IV.

\section{Non-rotating Boson-Fermion Stars}

The study of composed boson-fermion stars has been first addressed by
Henriques {\em et al} \cite{Henriques89}. In this section we present a
summary for boson stars model and its enhance for boson-fermion stars by
introducing fermions with Chandrasekhar perfect fluid model.

\subsection{Boson Stars}

A non-rotating configuration is spherically symmetric and one can build the
metric into the form:

\begin{equation}
ds^{2}=-B(r)d\tau ^{2}+A(r)dr^{2}+r^{2}d\theta ^{2}+r^{2}\sin ^{2}\theta
d\varphi ^{2}  \label{II.1}
\end{equation}
Our sign and conventions are the same as those used by Liddle and Madsen 
\cite{Liddle92}. The Lagrangian density for the complex scalar field with
mass {\em m}, is: 
\begin{equation}
{\cal L}=\frac{R}{16\pi G}-\partial _{\mu }\Phi ^{*}\partial ^{\mu }\Phi
-m^{2}\Phi ^{*}\Phi \,\,  \label{II.2}
\end{equation}
and, as usual for boson stars, we use the {\em ansatz}: 
\begin{equation}
\Phi (r,\tau )=\phi (r)e^{-i\omega \tau }\,\,  \label{II.3}
\end{equation}

The complex field $\phi (r)$ may be written as a sum of real and imaginary
parts. In absence of interaction between bosons and fermions, real and
imaginary parts generate similar sets of equations. This way one can
consider $\phi (r)$ purely real; this is not possible if fermions and bosons
interact \cite{Sousa98} since $\phi _{1}(r)$ and $\phi _{2}(r)$ present
different behaviour (considering $\phi (r)=\phi _{1}(r)+i\phi _{2}(r)$
above).

For simplicity we are not considering scalar field self-interaction term, $%
\lambda \Phi ^{4}$, and we are not considering boson-fermion interaction
although it should give more realistic results \cite{Colpi86}.
This way, Klein-Gordon equation has no source term: 
\begin{equation}
\Box \Phi -m^{2}\Phi =0  \label{II.3a}
\end{equation}

Equation (\ref{II.2}) gives the scalar-field (bosons) energy-momentum tensor:

\begin{equation}
T_{\mu \nu }^{B}=\partial _{\nu }\Phi ^{*}\partial _{\mu }\Phi +\partial
_{\mu }\Phi ^{*}\partial _{\nu }\Phi -g_{\mu \nu }(\partial _{\lambda }\Phi
^{*}\partial ^{\lambda }\Phi +m^{2}{\Phi }^{*}\Phi )  \label{II.4}
\end{equation}

From the N\"{o}ther theorem the action derived from (\ref{II.2}) gives rise
to a conserved current: 
\begin{equation}
J^{\mu }=ig^{\mu \nu }\left( \Phi ^{*}\partial _{\nu }\Phi -\Phi \partial
_{\nu }\Phi ^{*}\right)  \label{II.5}
\end{equation}
and the conserved charge:

\begin{equation}
N_{{\rm B}}=\int d^{3}r\sqrt{-g}J^{0}\,\,  \label{II.6}
\end{equation}
which can be identified with the number of bosons.

Once the scalar field describes particles a quantum analysis is required. If
we take $N_{B}$ bosons in their lowest state, $\mid N_{B}\rangle $, the
Einstein equations $G_{\mu \nu }=\langle N_{B}\mid :T_{\mu \nu
:}|N_{B}\rangle $ give exactly the same set of equations. This result is
obtained with the scalar field $\Phi $ changed into the operator: 
\begin{equation}
\Phi (r,t)=\sum\limits_{{\bf k}}\frac{1}{\sqrt{2\omega _{{\bf k}}}}\left[ a_{%
{\bf k}}\phi _{k}(r)e^{-i\omega _{{\bf k}}\tau }+a_{{\bf k}}^{\dagger }\phi
_{k}^{\dagger }(r)e^{i\omega _{{\bf k}}\tau }\right]  \label{II.7}
\end{equation}
where $a_{{\bf k}}^{\dagger }$ creates quanta associated with the field $%
\Phi $ and $a_{{\bf k}}$ destroys them, obeying the commutation relation $%
\left[ a_{i},a_{j}^{\dagger }\right] =\delta _{ij}$ . Considering the
ground-state, one can introduce a semiclassical scalar field:

\begin{equation}
\Phi _{{\rm c}}=\sqrt{\frac{N_{{\rm B}}+1/2}{\omega _{0}}}\phi
_{0}(r)e^{-i\omega _{0}\tau }  \label{II.8}
\end{equation}
where $\phi _{0}$ and $\omega _{0}$ are the eingenfunction of the
corresponding lowest eingenenergy, which gives a nodeless eingenfunction $%
\phi _{0}(r)$. Thence, forthcoming equations and numerical results are
related to the semiclassical field $\Phi _{{\rm c}}$.

The requirement of the functions to be orthonormal: 
\begin{equation}
\int_{0}^{\infty }4\pi r^{2}\sqrt{\frac{A}{B}}\phi _{i}^{\dagger }\phi
_{j}dr=\delta _{ij}  \label{II.10}
\end{equation}
allows one to determine the number of bosons, $N_{{\rm B}}$, by using the
classical field (\ref{II.8}): 
\begin{equation}
\int_{0}^{\infty }4\pi r^{2}\sqrt{\frac{A}{B}}\left| \Phi _{{\rm c}}\right|
^{2}dr=\frac{N_{{\rm B}}+1/2}{\omega _{0}}  \label{II.11}
\end{equation}

\subsection{Introducing Fermions}

Boson-fermion model star has been originally proposed by Henriques 
{\em et al} \cite{Henriques89}, and are equilibrium compact cold 
configurations made from both bosons and fermions.

As proposed by Chandrasekhar \cite{Chandrasekhar39} for neutron star models,
an equation of state can describe a perfect fluid of degenerate Fermi gas,
and density of energy and pressure are given by:

\[
\rho =K(\sinh t-t) 
\]

\begin{equation}  \label{II.12}
\end{equation}

\[
p=\frac{K}{3}\left( \sinh t-8\sinh \frac{t}{2}-3t\right) 
\]
where $K=m_{n}^{4}/32\pi ^{2}$, and $m_{n}$ is the fermion mass (which is
called neutron mass for neutron stars). The parameter $t$ is related to the
maximum momentum value, $q_{0}=q_{0}(r)$, in the Fermi distribution in a
distance $r$ from the centre: 
\begin{equation}
t=\ln \left\{ \frac{q_{0}}{m_{n}}+\sqrt{1+\left( \frac{q_{0}}{m_{n}}\right)
^{2}}\right\}  \label{II.13}
\end{equation}

The density of fermion particles in the distribution is:

\begin{equation}
n(r)=\frac{q_{0}^{3}}{3\pi ^{2}}=\frac{m_{n}^{3}}{3\pi ^{2}}\sinh {}^{3}(%
\frac{t}{4})  \label{II.14}
\end{equation}

Considering a compact spherically symmetric distribution with (\ref{II.1})
and (\ref{II.14}), the number of fermions is:

\begin{equation}
N_{{\rm F}}=\int 4\pi r^{2}n\sqrt{A}dr  \label{II.15}
\end{equation}

Oppenheimer and Volkoff \cite{Oppenheimer39} have found equilibrium
configurations by using Einstein equations with $%
T_{1}^{1}=T_{2}^{2}=T_{3}^{3}=p$ and $T_{0}^{0}=-\rho $, and with the
hydrostatic equilibrium equation for the pressure:

\begin{equation}
p^{\prime }=-\frac{(\rho +p)}{2}\frac{B^{\prime }}{B}  \label{II.16}
\end{equation}
where $^{\prime }=d/dr$, and $B=B(r)$ is defined on (\ref{II.1}).

As proposed by Henriques {\em et al} \cite{Henriques89}, the basic
properties of the composed boson-fermion star can be studied with the use of
an energy momentum tensor given by the sum of bosonic (B) and fermionic (F)
counterparts:

\begin{equation}
T_{\mu \nu }=T_{\mu \nu }^{{\rm B}}+T_{\mu \nu }^{{\rm F}}  \label{II.17}
\end{equation}
where: 
\begin{equation}
T_{\mu \nu }^{{\rm B}}=\partial _{\mu }\Phi \partial _{\nu }\Phi
^{*}+\partial _{\mu }\Phi ^{*}\partial _{\nu }\Phi -g_{\mu \nu }\left(
\partial _{\lambda }\Phi ^{*}\partial ^{\lambda }\Phi +m^{2}\Phi ^{*}\Phi +%
\frac{\lambda }{4}\Phi ^{4}\right)  \label{II.18}
\end{equation}
\begin{equation}
T_{\mu \nu }^{{\rm F}}=(\rho +p)u_{\mu }u_{\nu }+g_{\mu \nu }p  \label{II.19}
\end{equation}
where $u^{\mu }=(u^{0},u_{r},u_{\theta },u_{\varphi })=\frac{dx^{\mu }}{ds}$
is the fermion fluid four vector.

The set of differential equations can be obtained by using Einstein
equations together with (\ref{II.3a}) and (\ref{II.16}). After some
calculations and by using variables redefinitions:

\[
x\stackrel{.}{=}mr\,\,\,\,\,,\,\,\,\,\,\sigma (x)\stackrel{.}{=}\sqrt{8\pi G}%
\phi \,(r)\,\,\,\,,\,\,\,\,\,{\rm w}\stackrel{.}{=}\frac{\omega }{m} 
\]
\begin{equation}  \label{II.20}
\end{equation}
\[
\,\,\,\,\,\bar{\rho}(t)\stackrel{.}{=}\frac{4\pi G}{m^{2}}\rho
(t)\,\,\,\,\,,\,\,\,\,\,\bar{p}(t)\stackrel{.}{=}\frac{4\pi G}{m^{2}}p(t) 
\]
we obtain:

\begin{equation}
A^{\prime }=xA^{2}\left[ 2\bar{\rho}+\left( \frac{{\rm w}^{2}}{B}+1\right)
\sigma ^{2}+\frac{\sigma ^{\prime \,\,2}}{A}\right] -\frac{A}{x}(A-1)
\label{II.21}
\end{equation}
\begin{equation}
B^{\prime }=xAB\left[ 2\bar{p}+\left( \frac{{\rm w}^{2}}{B}-1\right) \sigma
^{2}+\frac{\sigma ^{\prime \,\,2}}{A}\right] +\frac{B}{x}(A-1)  \label{II.22}
\end{equation}
\begin{equation}
\sigma ^{\prime \prime }=-\left[ \frac{2}{x}+\frac{1}{2}\left( \frac{%
B^{\prime }}{B}-\frac{A^{\prime }}{A}\right) \right] \sigma ^{\prime
}-A\left[ \left( \frac{{\rm w}^{2}}{B}-1\right) \sigma -\Lambda \sigma
^{3}\right]  \label{II.23}
\end{equation}
\begin{equation}
t^{\prime }=-2\frac{B^{\prime }}{B}\frac{\sinh t-2\sinh (t/2)}{\cosh
t-4\cosh (t/2)+3}  \label{II.24}
\end{equation}

Also, any function $F(r)$ is replaced with $F(x/m)$. Hence, $F^{\prime }(x)=%
\frac{dF(x/m)}{dx}=\frac{1}{m}\frac{dF(r)}{dr}$, where now $^{\prime }=d/dx$%
. Equations (\ref{II.21}) - (\ref{II.22}) are obtained directly from
Einstein equations using metric (\ref{II.1}), and (\ref{II.23}) is the
Klein-Gordon equation (\ref{II.3a}). Equation (\ref{II.24}) arises from
hydrostatic equilibrium equation (\ref{II.16}), rewritten in the form: 
\begin{equation}
t^{\prime }=-\frac{1}{d\bar{p}/dt}\frac{B^{\prime }}{2B}(\bar{\rho}+\bar{p})
\label{II.25}
\end{equation}

where:

\begin{equation}
\bar{\rho}=\bar{K}(\sinh t-t)\,\,\,\,\,,\,\,\,\,\,\bar{p}=\frac{\bar{K}}{3}%
\left( \sinh t-8\sinh (t/2)+3t\right)  \label{II.26}
\end{equation}

with:

\begin{equation}
\bar{K}=\frac{m_{n}^{4}}{8\pi m^{2}M_{{\rm Pl}}^{2}}  \label{II.27}
\end{equation}

For convenience we will use $\bar{K}=1/4\pi $, which is the same value used
by Oppenheimer and Volkoff for neutron stars. Note that (\ref{II.27})
displays a constraint between boson mass $m$ and fermion mass $m_{n}$.
Hence, if one choose $m_{n}$ as the neutron mass and $\bar{K}=1/4\pi $, we
are setting $m=5,11\times 10^{-17}$MeV.

\section{Slowly Rotating Boson-Fermion Star}

Slowly rotating relativistic stars has been studied for long in applications
with neutron stars \cite{Hartle67}. Meanwhile, until Kobayashi {\em et al }%
\cite{Kobayashi94} rotation configurations for boson stars has never been
studied, and until Silveira {\em et al} \cite{Silveira95} stable boson stars
rotating configurations has never been found. Indeed, Kobayashi {\em et al}
show that there is no stable configuration with nonvanishing angular
momentum for pure boson star studied perturbatively. Meanwhile, considering
the quantum background for an axisymmetric scalar field one obtains stable
configurations in the Newtonian framework.

In this section we study relativistic equations for slowly rotating
boson-fermion stars, in a similar framework presented by Kobayashi {\em et al%
}.

\subsection{Rotating Fermion Stars}

Using Hartle's prescription \cite{Hartle67} for slowly rotating neutron
stars, one introduces the metric:

\begin{equation}
ds^{2}=-H^{2}d\tau ^{2}+Q^{2}dr^{2}+r^{2}K^{2}\left[ d\theta ^{2}+\sin
^{2}\theta \left( d\varphi -Ld\tau \right) ^{2}\right]  \label{III.1}
\end{equation}
where $H$, $Q$, $K$ and $L$ are functions of $r$ and $\theta $. When
considering slow rotation we have $R\Omega \ll c$, where $R$ is the average
radius and $\Omega $ is the angular velocity as seen by an observer at
infinity. In this section we expand the metric (\ref{III.1}) only up to
order $\Omega ^{2}$. This way, energy and pressure can be approximated using:

\[
{\cal E}=\rho +O(\Omega ^{2}) 
\]
\begin{equation}  \label{III.2}
\end{equation}
\[
{\cal P}=p+O(\Omega ^{2}) 
\]

When rotation is considered this metric introduces a dragging of inertial
frames. Thence, $L=d\varphi /d\tau $ is the angular velocity acquired by one
observer free falling the infinity to the point $(r,\theta )$.

The metric (\ref{III.1}) must be invariant under rotation reversion, {\em %
i.e.}, $\varphi \rightarrow -\varphi $ and $\Omega \rightarrow -\Omega $.
Thus, functions $H$, $Q$ and $K$ are even in powers of $\Omega $; meanwhile $%
L$ is odd. Since for slow rotation we study terms only up to order $\Omega
^{2}$, one can consider effects up to order $\Omega $ in $L(r,\theta )$,
represented by $C(r,\theta )$: 
\begin{equation}
L(r,\theta )=C(r,\theta )+O(\Omega ^{3})  \label{III.3}
\end{equation}

In cases when there is no rotation, $\Omega =0$ and the energy-momentum
tensor for the perfect fluid is: 
\begin{equation}
T_{\nu }^{\mu }=(\rho +p)u^{\mu }u_{\nu }+p\delta _{\nu }^{\mu }
\label{III.4}
\end{equation}

But, when $\Omega \neq 0$: 
\begin{equation}
T_{\nu }^{\mu }=({\cal E}+{\cal P})u^{\mu }u_{\nu }+P\delta _{\nu }^{\mu }
\label{III.5}
\end{equation}

In the case analysed by Hartle \cite{Hartle67}, the first order term in (\ref
{III.3}), $C(r,\theta )$, can be solved analytically, using the equation: 
\begin{equation}
R_{3}^{0}=8\pi GT_{3}^{0}  \label{III.6}
\end{equation}
we are considering $u^{\mu }=\left( u^{\tau },u^{r},u^{\theta },u^{\varphi
}\right) =\left( u^{1},0,0,u^{3}\right) $ due to symmetries involved. Since $%
d\varphi =\Omega d\tau $, one can use $u^{3}=\Omega u^{0}$. With the
normalization condiction $u^{\mu }u_{\mu }=-1$, one obtains $\left(
u^{0}\right) ^{2}\left[ g_{00}+2\Omega g_{03}+\Omega ^{2}g_{33}\right] =-1$.
Thus, the energy-momentum tensor component $T_{0}^{0}$ is: 
\begin{equation}
T_{3}^{0}=({\cal E}+{\cal P})u_{3}u^{0}=({\cal E}+{\cal P})\left[
(u^{0})^{2}g_{03}+u^{0}u^{3}g_{33}\right] =({\cal E}+{\cal P}%
)(u^{0})^{2}(g_{03}+\Omega g_{33})  \label{III.7}
\end{equation}

Expanding this term up to order $\Omega ^{2}$ yields:

\begin{equation}
T_{3}^{0}=(\rho +p)e^{-\nu }(\Omega -C)r^{2}\sin {}^{2}\theta +O(\Omega ^{3})
\label{III.8}
\end{equation}

Therefore, the field equation (\ref{III.6}) becomes:

\begin{equation}
\frac{1}{r^{4}}\partial _{r}\left[ r^{4}e^{-(\nu +\lambda )/2}\partial _{r}%
\overline{C}\right] +\frac{e^{(\lambda -\nu )/2}}{r^{2}\sin {}^{3}\theta }%
\partial _{\theta }\left( \sin {}^{3}\theta \partial _{\theta }\overline{C}%
\right) =16\pi G(\rho +p)e^{(\lambda -\nu )/2}\overline{C}  \label{III.9}
\end{equation}
where $\overline{C}=\Omega -C$. Defining the quantity: 
\begin{equation}
j(r)=e^{-(\nu +\lambda )/2}  \label{III.10}
\end{equation}
one obtains: 
\begin{equation}
\frac{dj}{dr}=-\frac{1}{2}\left( \nu ^{\prime }+\lambda ^{\prime }\right)
e^{-(\nu +\lambda )/2}  \label{III.11}
\end{equation}

From Einstein equations:

\[
\lambda ^{\prime }=8\pi Gre^{\lambda }\rho -\frac{1}{r}(e^{\lambda }-1) 
\]
\begin{equation}  \label{III.12}
\end{equation}
\[
\nu ^{\prime }=8\pi Gre^{\lambda }p+\frac{1}{r}(e^{\lambda }-1) 
\]
one is able to rewrite (\ref{III.11}) as: 
\begin{equation}
\frac{4}{r}\frac{dj}{dr}=-16\pi G(\rho +p)e^{(\lambda -\nu )/2}
\label{III.13}
\end{equation}

Thus (\ref{III.9}) becomes, up to first order in $\Omega $: 
\begin{equation}
\frac{1}{r^{4}}\partial _{r}\left[ r^{4}j(r)\partial _{r}\overline{C}\right] +%
\frac{4}{r}(\partial _{r}j)\overline{C}+\frac{e^{(\lambda -\nu )/2}}{r^{2}\sin
{}^{3}\theta }\partial _{\theta }\left( \sin {}^{3}\theta \partial _{\theta }%
\overline{C}\right) =0  \label{III.14}
\end{equation}

Expanding $\overline{C}$ in vector spherical harmonics: 
\begin{equation}
\overline{C}(r,\theta )=\sum_{l=1}^{\infty }\overline{C}_{l}(r)\left[ -\frac{1}{\sin
\theta }\partial _{\theta }P_{l}(\cos \theta )\right]  \label{III.15}
\end{equation}
and thus: 
\begin{equation}
\frac{1}{r^{4}}\partial _{r}\left[ r^{4}j(r)\partial _{r}\overline{C}_{l}\right] +%
\overline{C}_{l}\left[ \frac{4}{r}\partial _{r}j-e^{-(\lambda -\nu )/2}\frac{%
l(l+1)-2}{r^{2}}\right] =0  \label{III.16}
\end{equation}

When $r\rightarrow \infty $, $\lambda $ and $\nu $ vanish, and $j\rightarrow
1$. Thence, when $r\rightarrow \infty $ in (\ref{III.16}), $\overline{C_{l}}$
behaves like: 
\begin{equation}
\overline{C}_{l}(r)\rightarrow
k_{1}r^{-l-2}+k_{2}r^{l-1}\,\,\,\,\,\,\,\,\,\,\,\,\,\,\,\,\,\,\,\,(r%
\rightarrow \infty )  \label{III.17}
\end{equation}
where $k_{1}$ and $k_{2}$ are constants. Since fields decrease very fast the
solution when $r\rightarrow \infty $ is of the Kerr-Newman form: 
\[
C_{l}\propto \frac{1}{r^{3}} 
\]
in such way that$\ \overline{C}_{l}=\Omega -C_{l}\simeq \Omega .$ This is
reproduced by (\ref{III.17}) by choosing $l=1$, and appropriate values for $%
k_{1}$ and $k_{2}$. Taking again correct values for the constants in such
wise the solution to be regular, one can observe that terms with $l>1$
decrease even faster and $l=1$ mode can be considered as the dominant one.
Hence, $\overline{C}(r,\theta )\simeq \overline{C}_{1}(r)$, and $\overline{C}
$ becomes function of $r$ alone. Now, (\ref{III.16}) is: 
\begin{equation}
\frac{1}{r^{4}}\partial _{r}\left[ r^{4}j(r)\partial _{r}\overline{C}_{1}\right] +%
\frac{4}{r}(\partial _{r}j)\overline{C}_{1}=0  \label{III.18}
\end{equation}

Since the solution must be regular close to the origin, $\overline{C}_{1}(0)=$%
const., and considering the exterior solution, one obtains: 
\begin{equation}
\overline{C}_{1}(r)=\Omega -\frac{2J}{r^{3}}  \label{III.19}
\end{equation}
where $J$ is the total angular momentum of the fermion star. The quantity $%
J/\Omega $ defines its moment of inertia.

\subsection{Introducing Bosons}

After reviewing Hartle's perturbative approach for rotating fermion stars
(which are usually referred as neutron stars) it is convenient to introduce
bosons to obtain the composite boson-fermion star model, which is the
original part of this article.

Following the perturbative approach used by Hartle, we expand the metric up
to terms in $\Omega ^{2}$, and the Kerr metric with no charge can be written
as: 
\begin{equation}
ds^{2}=-B(r)d\tau ^{2}+A(r)dr^{2}+r^{2}d\theta ^{2}+r^{2}\sin {}^{2}\theta
d\varphi ^{2}-2C(r)r^{2}\sin {}^{2}\theta d\tau d\varphi +O(\Omega ^{2})
\label{III.20}
\end{equation}
where $C(r)=C_{1}(r)$ in the previous section. The total energy-momentum
tensor is again: 
\begin{equation}
T_{\mu \nu }=T_{\mu \nu }^{{\rm B}}+T_{\mu \nu }^{{\rm F}}  \label{III.21}
\end{equation}
as in equation (\ref{II.17}). Up to order $\Omega ^{2}$, $\stackrel{.}{%
\varphi }=\Omega $, and the nonvanishing components of $T_{\mu \nu }$ are: 
\begin{equation}
T_{0}^{0}=-\rho -\left( \frac{\omega ^{2}}{B}+m^{2}\right) \phi ^{2}-\frac{%
\phi ^{\prime \,\,2}}{A}  \label{III.22a}
\end{equation}
\begin{equation}
T_{1}^{1}=p+\left( \frac{\omega ^{2}}{B}-m^{2}\right) \phi ^{2}+\frac{\phi
^{\prime \,\,2}}{A}  \label{III.22b}
\end{equation}
\begin{equation}
T_{2}^{2}=T_{3}^{3}=p+\left( \frac{\omega ^{2}}{B}-m^{2}\right) \phi ^{2}-%
\frac{\phi ^{\prime \,\,2}}{A}  \label{III.22c}
\end{equation}
\begin{equation}
T_{3}^{0}=\frac{(\rho +p)}{B}r^{2}\sin {}^{2}\theta \left[ \Omega -C\right]
\label{III.22d}
\end{equation}
\begin{equation}
T_{0}^{3}=(\rho +p)\Omega  \label{III.22e}
\end{equation}

Using redefinitions (\ref{II.20}) and after some straightforward
calculations, one obtains up to order $\Omega ^{2}$, the set of differential
equations: 
\begin{equation}
A^{\prime }=xA^{2}\left[ 2\bar{\rho}+\left( \frac{{\rm w}^{2}}{B}+1\right)
\sigma ^{2}+\frac{\sigma ^{\prime \,\,2}}{A}\right] -\frac{A}{x}(A-1)
\label{III.24}
\end{equation}
\begin{equation}
B^{\prime }=xAB\left[ 2\bar{p}+\left( \frac{{\rm w}^{2}}{B}-1\right) \sigma
^{2}+\frac{\sigma ^{\prime \,\,2}}{A}\right] +\frac{B}{x}(A-1)
\label{III.25}
\end{equation}
\begin{equation}
\sigma ^{\prime \prime }=-\left[ \frac{2}{x}+\frac{1}{2}\left( \frac{%
B^{\prime }}{B}-\frac{A^{\prime }}{A}\right) \right] \sigma ^{\prime
}-A\left( \frac{{\rm w}^{2}}{B}-1\right) \sigma  \label{III.26}
\end{equation}
\begin{equation}
t^{\prime }=-2\frac{B^{\prime }}{B}\frac{\sinh t-2\sinh (t/2)}{\cosh
t-4\cosh (t/2)+3}  \label{III.27}
\end{equation}
\begin{equation}
C^{\prime \prime }=-\left[ \frac{4}{x}-\frac{1}{2}\left( \frac{B^{\prime }}{B%
}+\frac{A^{\prime }}{A}\right) \right] C^{\prime }-4A(\bar{\rho}+\bar{p}%
)\left( \Omega -C(r)\right)  \label{III.28}
\end{equation}
where now $^{\prime }=d/dx$.

Equations (\ref{III.24}) to (\ref{III.27}) are the same as those obtained
for boson-fermion stars. Equation (\ref{III.28}) gives the function $C(x)$,
corresponding to rotation of the boson-fermion star. Note that $C(x)$
depends not only on $x$, but the parameter $t$ and metric fields also
contributes to the rotation. Thus, step-by-step results that numerically
arise from (\ref{III.24}) - (\ref{III.27}) can modify the rotational term $%
C(x)$, via changes on the initial values, $\sigma _{0}$ and $t_{0}$, which
represents bosons and fermions contributions respectively. On the other
hand, $C(x)$ does not appear on equations (\ref{III.24}) - (\ref{III.27}).
Hence, changes on the initial value of $C(x)$ does not produce changes on $%
\sigma (x)$, $t(x)$, $A(x)\,$or $B(x)$, up to order $\Omega ^{2}$.

\section{Results}

After obtaining perturbatively, up to order $\Omega ^{2}$, the equations
that govern the rotation of a boson-fermion star, the results from section
3.2 are obtained numerically. Before that, equations (\ref{III.26}) and (\ref
{III.28})\ ought be rewritten to form a first order set of differential
coupled equations. Hence, equation (\ref{III.26}) splits into: 
\begin{equation}
s=\sigma ^{\prime } \label{III.29a}
\end{equation}

\begin{equation}
s^{\prime }=\left( xA\sigma ^{2}-\frac{A+1}{x}\right) s-A\sigma \left( \frac{%
{\rm w}^{2}}{B}-1\right) \label{III.29b}
\end{equation}
and taking: 
\begin{equation}
\overline{C}(x)=\Omega -C(x) \label{III.29c}
\end{equation}
equation (\ref{III.28}) becomes:
\begin{equation}
\overline{C}^{\prime }=-U \label{III.29d}
\end{equation}
\begin{equation}
U^{\prime }=\left[ xA\left( \bar{\rho}+\bar{p}+\frac{{\rm w}^{2}}{B}\sigma
^{2}+\frac{s^{2}}{A}\right) -\frac{4}{x}\right] U-4A\left( \bar{\rho}+\bar{p}%
\right) \overline{C} \label{III.29e}
\end{equation}

The boundary condictions are $A_{0}=1$, $s_{0}=0$, $U_{0}=U(0)=0$,
considering $B_{0}$, $\sigma _{0}$, $t_{0}$ and $\overline{C}(0)=\overline{C}%
_{0}$ as arbitrary initial values to be ascribed. And, at the infinity
condictions are $A(\infty )\rightarrow 1$, $B(\infty )\rightarrow 1$, $%
\sigma (\infty )\rightarrow 0$ and $s(\infty )\rightarrow 0$. As seem
before, star properties as mass, radius and total number of particles does
not differ from boson-fermion stars with no rotation up to $\Omega ^{2}$.
For example, a configuration with $\sigma _{0}=0.2$, $\overline{C}_{0}=1$
and $t_{0}=4.0$ we have obtained $M\simeq 0.41M_{{\rm Pl}}^{2}/m$, $(mN_{%
{\rm B}}+m_{n}N_{{\rm F}})\simeq 0.46M_{{\rm Pl}}^{2}/m$ e $mR\simeq 0.94M_{%
{\rm Pl}}^{2}/m$. Thus, if one finds an equilibrium configuration, changes on
rotation parameter $\overline{C}$ does not modify those results. Meanwhile,
if one fix the initial value $\overline{C}_{0}$, changes on values of $%
\sigma _{0}$ and $t_{0}$ provide modifications on the $\overline{C}$ values
evolution, {\em i.e.}, the referential frames dragging is modified.

Evolution of the scalar fields, its derivative and metric coefficients are
shown in figure 1, and rotation parameters are shown in figure 2. Those
figures are obtained with $\sigma _{0}=0.20$, $t_{0}=4.0$ and 
$\overline{C}_{0}=1.0$,
for which $B_{0}=0.1422$ and ${\rm w}=0.79108$. In this case, $E_{{\rm B}}<0$%
, showing that configuration is probably stable. We can also observe that $%
\overline{C}$ stabilizes for $x\rightarrow \infty $. But this is expected
and from equation (\ref{III.19}) one can see that this maximum value is $%
\Omega $, up to the approximation order we assume. 
One can obtain this by using the boundary condictions at infinity, where 
asymptotic flatness is expected for (\ref{III.1}), and 
$\lim_{r\rightarrow\infty}A(r)=1=\lim_{r\rightarrow\infty}B(r)$ and
$\lim_{r\rightarrow\infty}C(r)=0$. The last one is equivalent to:
\begin{equation}
\lim_{x\rightarrow\infty}C(x)=0 \label{III.30}
\end{equation}

Since $C(x)=\Omega - \overline{C}(x)$, equation (\ref{III.30}) is 
possible if and only if:
\begin{equation}
\lim_{x\rightarrow\infty}\overline{C}(x)=\Omega \label{III.31}
\end{equation}

In the case shown in the figures 1 and 2, angular momentum  as seem 
by an observer far from the object is $\Omega = -0.1651$. Some other 
results are shown in the table~I below for a configuration with
$t_0 =4.0$ and $\sigma_0 =0.2$.

\begin{center}
\begin{table}
\begin{tabular}{|c||c|c|c|c|c|} 
{\bf $\overline{C}_0$}& 3.0 & 2.0 & 1.0 & -1.0 & -2.0  \\ \hline
{\bf $\Omega$}& -0.4954 & -0.3302 & -0.1651 & 0.1651 & 0.3302 \\
\end{tabular}
  \caption{Values obtained for $\Omega$ given an initial value 
of the rotational parameter $\overline{C}_0$, with 
$\sigma_0 = 0.2$ e $t_0 =4.0$.}
\end{table}
\end{center}

All configurations obtained present $E_{\rm B}<0$, and the same values
for mass and radius. Note that negative values just provide an inversion
of rotation, as expected.


\begin{figure}
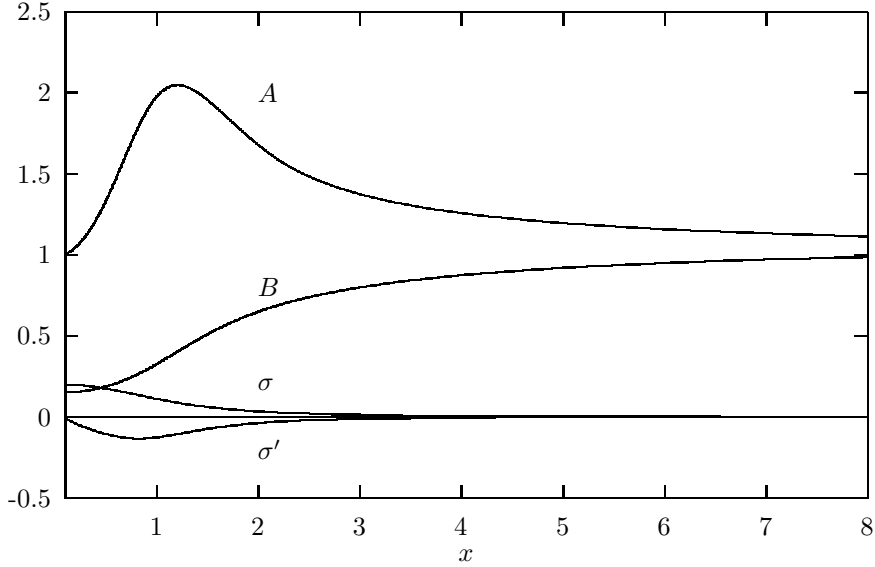

\setlength{\unitlength}{0.240900pt}
\ifx\plotpoint\undefined\newsavebox{\plotpoint}\fi
%
   \caption{Scalar field, its derivative and metric coeffcients for a
composite boson-fermion star,\thinspace with 
$\sigma _{0}=0.20$, $t_{0}=4.0$ and $C_{0}=1.0$. Values obtained: 
$B_{0}=0.1422$, ${\rm w}=0.79108$ and $E_{{\rm B}}\simeq -0.05$. }
\end{figure}

\begin{figure}
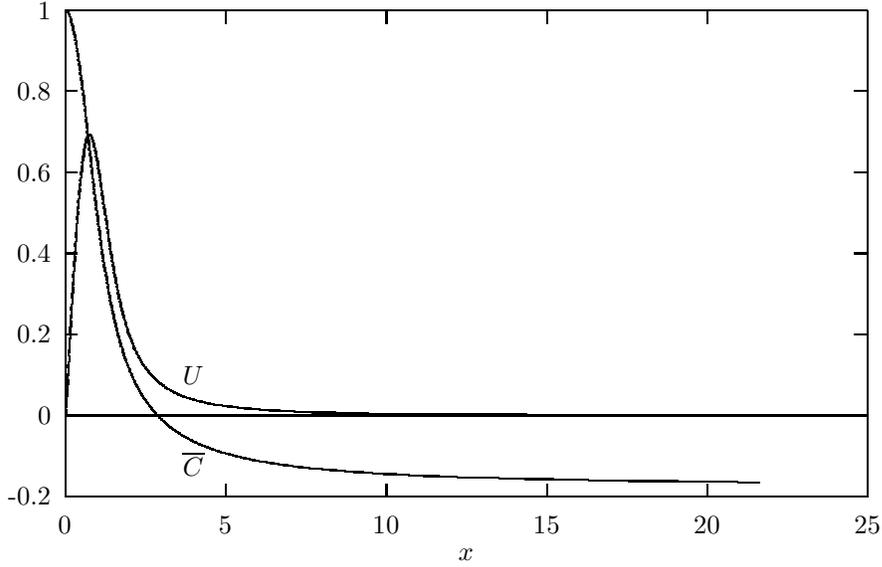

\setlength{\unitlength}{0.240900pt}
\ifx\plotpoint\undefined\newsavebox{\plotpoint}\fi
%
 \caption{Rotation parameter evolution for the case shown in figure 1. In 
 this case $\lim_{x\rightarrow\infty}\overline{C}(x)=\Omega =-0.1651$.}
\end{figure}

\section{Conclusions}

Given a star composed of both bosons and fermions, without interaction, we
have show that it is possible to obtain stable configurations with slow
rotation, by using the perturbative relativistic method that usually
describes neutron stars. We have developed sample calculations considering
fields up to the second order in angular velocity. With that, angular
velocity, mass, number of particles and binding energy can be measured by an
observer at the infinity.

But, if it is known \cite{Kobayashi94} that the perturbative prescription
allows no rotation for boson stars, why have we obtained rotation with the
same prescription when bosons are mixed to fermions inside the star?\ The
answer is in equation (\ref{III.28}). Up to the order considered here,
rotation is not directly influenced by the scalar field; instead, energy
density and pressure of the fermi gas are explicit in the equation that
describes the rotation of the configuration.

Since we have performed $l=0$ case, which is the predominant one, these
results may contribute to studies on deformations of the structure, {\em e.g.%
} $l=2$, and so on, and studies on the emission of gravitational waves from
those kind of objects. This study also opens some questions about spinning
boson fermion stars with the use of axisymmetric scalar field.


\begin{references}
\bibitem{Oppenheimer39}  J.\ R.\ Oppenheimer \&\ G.\ M.\ Volkoff, {\em On
Massive Neutron Cores}, Phys. Rev. {\bf 55}, 374-364\ (1939); Some authors 
also cite: W. Baade \& F. Zwicky {\em On Supernovae}, Proc. Nat. Acad. Sci.
(USA) {\bf 20}, 245-259 (1934).

\bibitem{Ruffini69}  R. Ruffini and S. Bonazzola, {\em Systems of
Self-Gravitating Particles in General Relativity and the Concept of an
Equation of State}, Phys. Rev. {\bf 187}, 1767-1783(1969); D. J. Kaup, {\em %
Klein-Gordon Geons}, Phys. Rev. {\bf 172}, 1331-1342(1968).

\bibitem{Breit84}   J. D. Breit, S. Gupta and A. Zaks, {\em Cold Boson Stars}%
, Phys. Lett. {\bf B 140}, 329-332 (1984).

\bibitem{Liddle92}  A. R. Liddle and M. S. Madsen, {\em The Structure and
Formation of Boson Stars}, Int. J. Mod. Phys.{\bf \ D1}, 101-143(1992); Ph.
Jetzer, {\em Boson Stars}, Phys. Rep. {\bf 220}, No.4, 163-227 (1992); E. W.
Mielke and F. E. Schunk, {\em Boson Stars:\ Early History and Recent
Prospects}, gr-qc/9801063 (1998).

\bibitem{Henriques89}  A.\ B.\ Henriques, A.\ R.\ Liddle \&\ R.\ G.\
Moorhouse, {\em Combined Boson-Fermion Stars}, Phys. Lett. {\bf B233},
99-106 (1989).

\bibitem{Chandrasekhar39}  S.\ Chandrasekhar, {\em An Introduction to the
Study of Stellar Structure} (Dover, 1939).

\bibitem{Gleiser88}  M.\ Gleiser, {\em Stability of Boson Stars}, Phys. Rev. 
{\bf D38}, 2376-2385 (1988); M.\ Gleiser \& R. Watkins, {\em Gravitational
Stability of Scalar Matter}, Nucl. Phys. {\bf B319}, 733-746 (1989).

\bibitem{Henriques90}  A.\ B.\ Henriques {\em et al}, {\em Combined
Boson-Fermion Stars: Configurations and Stability}, Nucl. Phys. {\bf B337},
737-761 (1990); A.\ B.\ Henriques {\em et al}, {\em Stability of
Boson-Fermion Stars}, Phys. Lett. {\bf B251}, 511-516 (1990).

\bibitem{Torres97}  D. F. Torres {\em Boson Stars in General Scalar-Tensor
Gravitation: Equilibrium Configurations}, Phys. Rev. {\bf D56}, 3478-3484
(1997).

\bibitem{Hartle67}  J.\ B.\ Hartle, {\em Slowly Rotating Relativistic Stars
I. Equations of Structure}, Ap. J. {\bf 150}, 1005-1029 (1967); J.\ B.\
Hartle \& K.\ S.\ Thorne, {\em Slowly Rotating Relativistic Stars II. Models
for Neutron Stars and Supermassive Stars}, Ap. J. {\bf 153}, 807-834 (1968).

\bibitem{Kobayashi94}  Y. Kobayashi, M. Kasai \& T. Futamase, {\em Does a
Boson Star Rotate?}, Phys. Rev. {\bf D50}, 7721-7724 (1994).

\bibitem{Silveira95}  V. Silveira \& C. M. G. de Sousa, {\em Boson Star
Rotation: a Newtonian Approximation}, Phys. Rev. {\bf D52}, 5724-5728 (1995).

\bibitem{Ryan97}  F. D. Ryan, {\em Spinning Boson Stars with Large
Self-Interaction}, Phys. Rev. {\bf D55} (1997); K. Sakamoto \& K. Shiraishi, 
{\em Rotating Boson Stars with Large Self-Interaction in (2+1) Dimensions},
gr-qc/9910113 (1999).

\bibitem{Yoshida97}  S. Yoshida \& \ Y. Eriguchi, {\em Rotating Boson Stars
in General Relativity}, Phys. Rev. {\bf D56}, 762-771 (1997); S. Yoshida \&
\ Y. Eriguchi, {\em New Static Axisymmetric and Nonvacuum Solutions in
General Relativity:\ Equilibrium Solutions of Boson Stars}, Phys. Rev. {\bf %
D55}, 1994-2001 (1997); S. Yoshida \& \ Y. Eriguchi, {\em Nonaxisymmetric
Boson Stars in Newtonian Gravity}, Phys. Rev. {\bf D56,} 1994-2001 (1997).

\bibitem{Sousa98}  C.\ M.\ G.\ de Sousa, J.\ L.\ Tomazelli and V. Silveira, 
{\em Model for Stars of Interacting Bosons and Fermions}, Phys. Rev. {\bf D58%
}, 123003 (1998).

\bibitem{Colpi86}  M. Colpi, S.\ L. Shapiro \& I. Wasserman, {\em Boson
Stars:\ Gravitational Equilibria of Self-Interacting Scalar Fields}, Phys.
rev. Lett. {\bf 57}, 2485-2488 (1986).

\bibitem{Grasso90}  D. Grasso, {\em Boson-Star Formation by Classical
Instability}, Phys. Rev. {\bf D41}, 2998-3002 (1990).
\end{references}
\end{document}